\documentclass[sigconf]{acmart}

\AtBeginDocument{%
  }

\usepackage{amsmath}
\usepackage{booktabs}
\usepackage{caption}
\usepackage{subcaption}
\usepackage{multirow}
\usepackage{diagbox}
\usepackage{colortbl}
\usepackage{algorithm}
\usepackage{adjustbox}
\usepackage{algpseudocode}
\usepackage{graphicx}
\usepackage{dirtytalk}
\usepackage{comment}

\copyrightyear{2026}
\acmYear{2026}
\setcopyright{cc}
\setcctype{by}
\acmConference[CIKM '26]{Proceedings of the 35th ACM International Conference on Information and Knowledge Management}{November 07--11, 2026}{Rome, Italy}
\acmBooktitle{Proceedings of the 35th ACM International Conference on Information and Knowledge Management (CIKM '26), November 07--11, 2026, Rome, Italy}
\acmDOI{10.1145/3799682.3840002}
\acmISBN{979-8-4007-2539-5/2026/11}

\begin{document}


\title{Reducing Representational Interference in Retrieval Embeddings with Trainable Prefix Tokens}


\author{Seongtae Hong}
\orcid{0009-0002-2073-7731}
\affiliation{%
  \institution{Korea University}
  \city{Seoul}
  \country{South Korea}
}
\email{ghdchlwls123@korea.ac.kr}

\author{Youngjoon Jang}
\orcid{0009-0005-3591-638X}
\affiliation{%
  \institution{Korea University}
  \city{Seoul}
  \country{South Korea}
}
\email{dew1701@korea.ac.kr}

\author{Jungseob Lee}
\orcid{0000-0002-9431-6342}
\affiliation{%
  \institution{Korea University}
  \city{Seoul}
  \country{South Korea}}
\email{omanma1928@korea.ac.kr}

\author{Chanjun Park}
\orcid{0000-0002-7200-9632}
\authornotemark[1]
\affiliation{%
  \institution{Soongsil University}
  \city{Seoul}
  \country{South Korea}}
\email{chanjun.park@ssu.ac.kr}

\author{Heuiseok Lim}
\orcid{0000-0002-9269-1157}
\authornote{Corresponding author}
\affiliation{%
  \institution{Korea University}
  \city{Seoul}
  \country{South Korea}}
\email{limhseok@korea.ac.kr}

\renewcommand{\shortauthors}{Seongtae Hong, Youngjoon Jang, Jungseob Lee, Chanjun Park, and Heuiseok Lim}

\begin{abstract}
Embedding models compress variable-length text into fixed-size vector representations, but this compression can entangle semantic content with residual token-level information, such as artifacts arising from attention sink behavior. We propose a simple method that prepends trainable \texttt{[DUMP]} tokens to the input during embedding model training. The method keeps the model architecture, pooling strategy, and training loss unchanged, while optimizing the added tokens through the standard contrastive retrieval objective. By introducing trainable prefix positions that interact with regular tokens during encoding, \texttt{[DUMP]} tokens help reduce the influence of residual or uninformative token-level signals on the final retrieval representation. Experiments on standard retrieval and long-context benchmarks show that our method consistently improves over finetuned baselines across encoder and decoder embedding models. Further analysis shows that \texttt{[DUMP]} tokens alter token-level activation and attention patterns associated with low-information tokens. These results suggest that trainable prefix tokens provide a lightweight way to improve retrieval representations.
\end{abstract}



\maketitle
\section{Introduction}
Embedding models have become a fundamental component of modern retrieval systems, where queries and passages are mapped into a shared vector space for efficient semantic matching~\cite{dpr,sentenceembedding}. Their effectiveness heavily depends on the ability to compress variable-length text into fixed-dimensional representations while preserving the critical semantic information relevant to retrieval across diverse domains. This compression is typically performed through pooling or token-selection strategies~\cite{pooling1,pooling2}, effectively forcing the model to funnel rich contextual dynamics into a narrow informational bottleneck. However, this process can also introduce representational interference: token-level information that is not directly useful for semantic matching, such as recurrent positional patterns, special-token artifacts, or other non-semantic cues, may be absorbed into the embedding representation and distort its true semantic content~\cite{streaming,register,quantifying}. When this noise accumulates, it fundamentally limits the discriminative power of the resulting embeddings in downstream retrieval tasks.

This issue becomes particularly important in light of the attention sink phenomenon, where certain positions attract disproportionately large attention despite carrying limited semantic content~\cite{streaming,zhang2025attention}. Since the final retrieval representation is obtained by pooling or selecting from contextualized token representations, such artificial attention bias can cause non-semantic information to become deeply entangled with content-bearing tokens~\cite{jtcse,lmk}, thereby diluting the overall sequence representation. Standard retrieval fine-tuning addresses this issue only indirectly: contrastive objectives improve query--passage alignment at the macro embedding level, but they do not explicitly control how residual token-level information is represented or routed before pooling. Consequently, conventional retrieval optimization may still leave residual token-level information entangled with the embedding representation, leaving a gap in how we optimize dense retrievers.

In this work, we focus on how such residual information is handled during retrieval encoding by proposing a simple yet highly effective architectural intervention. We use trainable \texttt{[DUMP]} tokens as input prefixes during embedding model training. These prefix tokens participate in contextualization before the sequence is compressed into a single embedding, acting as dedicated semantic buffers. By absorbing attention associated with uninformative signals, they help reduce the influence of residual token-level information that would otherwise be mixed into the embedding representation. Crucially, the method does not require modifying the core model architecture, introducing auxiliary losses, or changing the retrieval objective, making it a highly scalable plug-and-play solution. Experiments on six diverse embedding models show that our method yields average relative improvements of 2.46\% on standard retrieval benchmarks and 2.41\% under long-context retrieval settings, demonstrating robust generalization. Further analyses show that the number of prefix tokens affects performance in a non-monotonic manner, while token-level inspection indicates that the added prefixes successfully change attention patterns associated with residual or uninformative tokens, effectively routing noise away from the final sequence representation.

\section{RELATED WORK}

Dense retrieval models encode queries and documents into fixed-dimensional representations and compare them in a shared embedding space to enable highly efficient semantic matching. In practice, these models typically obtain sequence-level embeddings through pooling or token-selection strategies, such as \texttt{[CLS]}, mean, or last-token pooling~\cite{llm2vec,pooling2,lmk}. Although these strategies provide a simple way to aggregate contextualized token representations, they also create a severe compression bottleneck, as heterogeneous token-level information is collapsed into a single retrieval vector, often causing fine-grained semantic details to be lost. Prior work has recognized this bottleneck in dense passage retrieval and proposed specialized pre-training strategies to improve the quality of the compressed representation~\cite{retromae,dupmae}. However, while these methods enhance the general representation space, they do not inherently resolve the internal routing of noisy or uninformative signals during the actual encoding phase.

Recent studies on Transformer attention dynamics further suggest that not all token-level information contributes equally to the final semantic representation. In particular, the attention sink phenomenon shows that certain tokens can attract disproportionately large attention scores despite carrying limited semantic content~\cite{streaming,attentionsink1,sun2024massive,zhang2025attention}. Driven by the softmax operation that forces attention allocation~\cite{softpick,attentionsink2}, such behavior can help stabilize autoregressive generation, but it inadvertently introduces representational interference in embedding models, where the pooled vector is strictly expected to reflect the pure semantic content of the input. Related observations have also been made in vision Transformers, where learnable register tokens are introduced to absorb high-norm outlier information and reduce non-semantic artifacts in feature maps~\cite{register}. These studies directly motivate our use of trainable \texttt{[DUMP]} prefix tokens to provide a dedicated structural pathway that absorbs residual or non-semantic information during retrieval encoding, yielding cleaner and more discriminative sequence representations.
\section{Methods}
We use \texttt{[DUMP]} tokens as trainable prefixes during embedding model training, aiming to reduce residual token-level information by providing dedicated slots for the model to route non-semantic attention. The model architecture and retrieval objective remain unchanged. We first describe the encoding process and then present the retrieval training objective.

\subsection{Dump-Token Conditioned Encoding}
We describe the encoding process of an embedding model and then explain how \texttt{[DUMP]} tokens are incorporated as input prefixes. Let $q$ and $p$ denote a query and a passage, respectively. For an input text $x \in \{q,p\}$, a standard embedding model maps $x$ into a fixed-dimensional representation through tokenization, model encoding, and pooling. We add special tokens, denoted as \texttt{[DUMP]}, to the tokenizer vocabulary. In practice, we use $K$ indexed \texttt{[DUMP]} tokens, i.e., $[\mathrm{DUMP}_1], \ldots, [\mathrm{DUMP}_K]$. The embeddings of these tokens are initialized in the model's embedding matrix and updated jointly with the model parameters during retrieval fine-tuning. For $x \in \{q,p\}$, we prepend \texttt{[DUMP]} tokens:

\begin{equation}
    \tilde{x} = [\mathrm{DUMP}_1] \cdots [\mathrm{DUMP}_K] \oplus x,
\end{equation}
where $\oplus$ denotes text concatenation. During the forward pass, these prefix tokens interact with the regular text tokens via self-attention, allowing them to absorb structural or uninformative noise. The resulting embedding is computed as:

\begin{equation}
    \widetilde{E}_\theta(x) = \operatorname{pooling}\left(M_\theta(T(\tilde{x}))\right),
\end{equation}
where $T(\cdot)$ denotes the tokenizer, $M_\theta$ denotes the embedding model, and $\operatorname{pooling}(\cdot)$ denotes the pooling strategy. Crucially, the pooling strategy still operates on its default target (e.g., the \texttt{[CLS]} or last token), while the contextualized representation is influenced by interactions with the \texttt{[DUMP]} prefixes, which are introduced to reduce representational interference. The same prefixed encoding is used during both fine-tuning and inference. The \texttt{[DUMP]} tokens receive no separate supervision or auxiliary objective, and their embeddings are updated only through the retrieval loss.

\begin{table*}[t!]
\centering
\scriptsize
\setlength{\tabcolsep}{3.2pt}
\renewcommand{\arraystretch}{1.1}
\caption{Retrieval results on standard benchmark datasets. The table reports nDCG@10 and compares fine-tuned baseline models with their counterparts trained using \texttt{[DUMP]} tokens across encoder and decoder architectures.}
\label{tab:model_comparison}
\resizebox{\textwidth}{!}{
\begin{tabular}{l *{7}{|cc}}
\toprule
\multirow{2}{*}{\textbf{Model}}
& \multicolumn{2}{c}{\textbf{Wikipedia}}
& \multicolumn{2}{c}{\textbf{NQ}}
& \multicolumn{2}{c}{\textbf{HotpotQA}}
& \multicolumn{2}{c}{\textbf{Quora}}
& \multicolumn{2}{c}{\textbf{DBPedia}}
& \multicolumn{2}{c}{\textbf{MSMARCO}}
& \multicolumn{2}{c}{\textbf{Avg.}} \\
\cmidrule(lr){2-3}
\cmidrule(lr){4-5}
\cmidrule(lr){6-7}
\cmidrule(lr){8-9}
\cmidrule(lr){10-11}
\cmidrule(lr){12-13}
\cmidrule(lr){14-15}
& \textbf{Base} & \textbf{Ours}
& \textbf{Base} & \textbf{Ours}
& \textbf{Base} & \textbf{Ours}
& \textbf{Base} & \textbf{Ours}
& \textbf{Base} & \textbf{Ours}
& \textbf{Base} & \textbf{Ours}
& \textbf{Base} & \textbf{Ours} \\
\midrule

\multicolumn{15}{c}{Encoder Architecture} \\
\midrule
gte-en-mlm-base
& 0.8348 & \textbf{0.8385}
& \textbf{0.3828} & 0.3793
& 0.5200 & \textbf{0.5268}
& 0.8266 & \textbf{0.8388}
& 0.3632 & \textbf{0.3655}
& 0.6258 & \textbf{0.6310}
& 0.5922 & \textbf{0.5966} \\

NeoBERT-base
& 0.8500 & \textbf{0.8534}
& \textbf{0.3944} & 0.3830
& \textbf{0.5469} & 0.5440
& 0.8279 & \textbf{0.8369}
& 0.3683 & \textbf{0.3719}
& 0.5772 & \textbf{0.5909}
& 0.5941 & \textbf{0.5967} \\

EuroBERT-210m
& 0.8328 & \textbf{0.8518}
& 0.3849 & \textbf{0.3881}
& 0.5240 & \textbf{0.5276}
& 0.8124 & \textbf{0.8270}
& 0.3016 & \textbf{0.3103}
& 0.6158 & \textbf{0.6256}
& 0.5786 & \textbf{0.5884} \\

bge-m3-retromae
& 0.8521 & \textbf{0.8648}
& 0.4062 & \textbf{0.4062}
& \textbf{0.5454} & 0.5430
& 0.8448 & \textbf{0.8512}
& 0.3409 & \textbf{0.3456}
& 0.6286 & \textbf{0.6287}
& 0.6030 & \textbf{0.6066} \\

\midrule
\multicolumn{15}{c}{Decoder Architecture} \\
\midrule
ettin-decoder-150m
& 0.7277 & \textbf{0.7601}
& 0.1839 & \textbf{0.2090}
& 0.3861 & \textbf{0.4215}
& 0.7135 & \textbf{0.7703}
& 0.1889 & \textbf{0.2160}
& 0.4264 & \textbf{0.4882}
& 0.4377 & \textbf{0.4775} \\

MobileLLM-R1-140M
& 0.8539 & \textbf{0.8620}
& 0.3779 & \textbf{0.4102}
& 0.5425 & \textbf{0.5572}
& 0.8251 & \textbf{0.8346}
& 0.3047 & \textbf{0.3478}
& 0.6121 & \textbf{0.6438}
& 0.5860 & \textbf{0.6093} \\

\bottomrule
\end{tabular}
}
\end{table*}

\subsection{Training Objective}
\label{obj}
We fine-tune the embedding model using retrieval pairs $\{(q_i,p_i^+)\}_{i=1}^{N}$, where $q_i$ denotes a query and $p_i^+$ denotes its positive passage. Before encoding, the \texttt{[DUMP]} tokens are prepended to both queries and passages. We define the embeddings and the scaled similarity function as:
\begin{equation}
    \mathbf{e}_x = \widetilde{E}_\theta(x), 
    \qquad 
    s(q,p) = \cos(\mathbf{e}_q, \mathbf{e}_p)/{\tau}.
\end{equation}
The model is optimized with the InfoNCE loss:

\begin{equation}
    \mathcal{L}_{\text{NCE}}
    =
    -\log
    \frac{
        e^{s(q_i,p_i^+)}
    }{
        e^{s(q_i,p_i^+)}
        +
        \sum_{j=1}^{n} e^{s(q_i,p_{i,j}^-)}
    },
\end{equation}
where $\tau$ is the temperature parameter and $\{p_{i,j}^-\}_{j=1}^{n}$ denotes negative passages for query $q_i$. This objective increases the similarity between each query and its positive passage while separating it from negative passages. Since the \texttt{[DUMP]} token embeddings are part of the trainable parameters, they are updated through the same retrieval loss without any auxiliary objective. Because the final pooled vector relies on the attention distributions formed across the entire sequence, the training signal from the contrastive loss naturally backpropagates into the \texttt{[DUMP]} token embeddings, optimizing them to better manage non-semantic cues.

\section{Experiment}
\subsection{Experimental Setup}

\noindent\textbf{Training.}
We implement our method across a diverse set of encoder and decoder architectures. For the encoder models, we utilize gte-en-mlm-base~\cite{gte}, NeoBERT-base~\cite{neobert}, EuroBERT-210m~\cite{eurobert}, and bge-m3-retromae~\cite{bge}. For the decoder models, we employ ettin-decoder-150m~\cite{ettin} and MobileLLM-R1-140M~\cite{mobile-r1}. For encoder models, we use \texttt{[CLS]} pooling, while for decoder models, we use last-token pooling~\cite{qwen3}. We set the number of \texttt{[DUMP]} tokens to $K=4$ in our main experiments, corresponding to \texttt{[DUMP1]}, \texttt{[DUMP2]}, \texttt{[DUMP3]}, and \texttt{[DUMP4]}. For training, we construct a dataset of 110k instances, utilizing one negative passage per query. Specifically, we use 100k examples from RLHN~\citep{rlhn} to leverage its relabeled hard negatives, which provide cleaner contrastive supervision by mitigating the influence of false negatives. To enhance long-context capabilities, we additionally include 10k instances from MLDR~\citep{bge}, a long-document retrieval dataset constructed from lengthy Wikipedia and mC4 articles. All models are fine-tuned for one epoch using a maximum sequence length of 8192, a learning rate of $1 \times 10^{-4}$, and a contrastive temperature scale of 20.0. Training is conducted with bfloat16 precision and gradient caching~\citep{gradcache}, maintaining a global batch size of 256 to gather sufficient cross-device in-batch negatives.

\noindent\textbf{Evaluation.}
We evaluate retrieval performance using the MTEB framework~\citep{mteb} on a subset of the BEIR benchmark~\citep{beir}. To rigorously test the models' discriminative capabilities, we specifically utilize the hard-negative versions of Natural Questions~\citep{nq}, HotpotQA~\citep{hotpotqa}, Quora~\citep{quora}, DBPedia~\citep{dbpedia}, and MS MARCO~\citep{msmarco}. We additionally evaluate on the English subset of Wikipedia from MTEB~\citep{mteb} to measure performance on a large-scale general-domain corpus. Furthermore, we use LongEmbed~\citep{longembed} to assess long-context embedding performance, where models must effectively compress substantially longer inputs into fixed representations. We report nDCG@10 for all retrieval evaluations.

\subsection{Results}

\noindent\textbf{Main Results.}
Table~\ref{tab:model_comparison} reports retrieval performance across different embedding models. The \textit{Base} models are fine-tuned with the standard retrieval objective without additional prefix tokens, whereas \textit{Ours} prepends trainable \texttt{[DUMP]} tokens during training. Overall, our method consistently improves the average performance across all evaluated models, indicating that adding trainable prefix tokens is broadly effective for retrieval embedding learning. The gains are consistent across both encoder and decoder architectures; for example, ettin-decoder improves from 0.4377 to 0.4775 on average, while EuroBERT improves from 0.5786 to 0.5884.

These consistent improvements support our hypothesis that assigning dedicated \texttt{[DUMP]} positions acts as a useful auxiliary buffer to reduce representational interference during query--passage encoding. Since the model architecture and retrieval objective remain unchanged, these results demonstrate that trainable prefix tokens provide a simple yet effective way to improve retrieval representations without requiring auxiliary losses or structural modifications.

\begin{table}[t]
\centering
\small
\caption{Long-context retrieval performance on LongEmbed, reported in nDCG@10. The table compares baseline models with models trained using \texttt{[DUMP]} tokens.}
\label{tab:longembed}
\renewcommand{\arraystretch}{1.1}
\resizebox{0.75\linewidth}{!}{
\begin{tabular}{llcc}
\toprule
\textbf{Arch.} & \textbf{Model} & \textbf{Base} & \textbf{Ours} \\
\midrule
\multirow{4}{*}{Encoder}
& gte-en-mlm-base      & 0.4797 & \textbf{0.5024} \\
& NeoBERT-base         & 0.3391 & \textbf{0.3452} \\
& EuroBERT-210m        & 0.4313 & \textbf{0.4363} \\
& bge-m3-retromae      & 0.5548 & \textbf{0.5679} \\
\midrule
\midrule
\multirow{2}{*}{Decoder}
& ettin-decoder-150m   & 0.3330 & \textbf{0.3498} \\
& MobileLLM-R1-140M    & 0.5580 & \textbf{0.5593} \\
\bottomrule
\end{tabular}
}
\end{table}

\begin{figure*}[t!]
    \centering
    \includegraphics[width=0.9\textwidth]{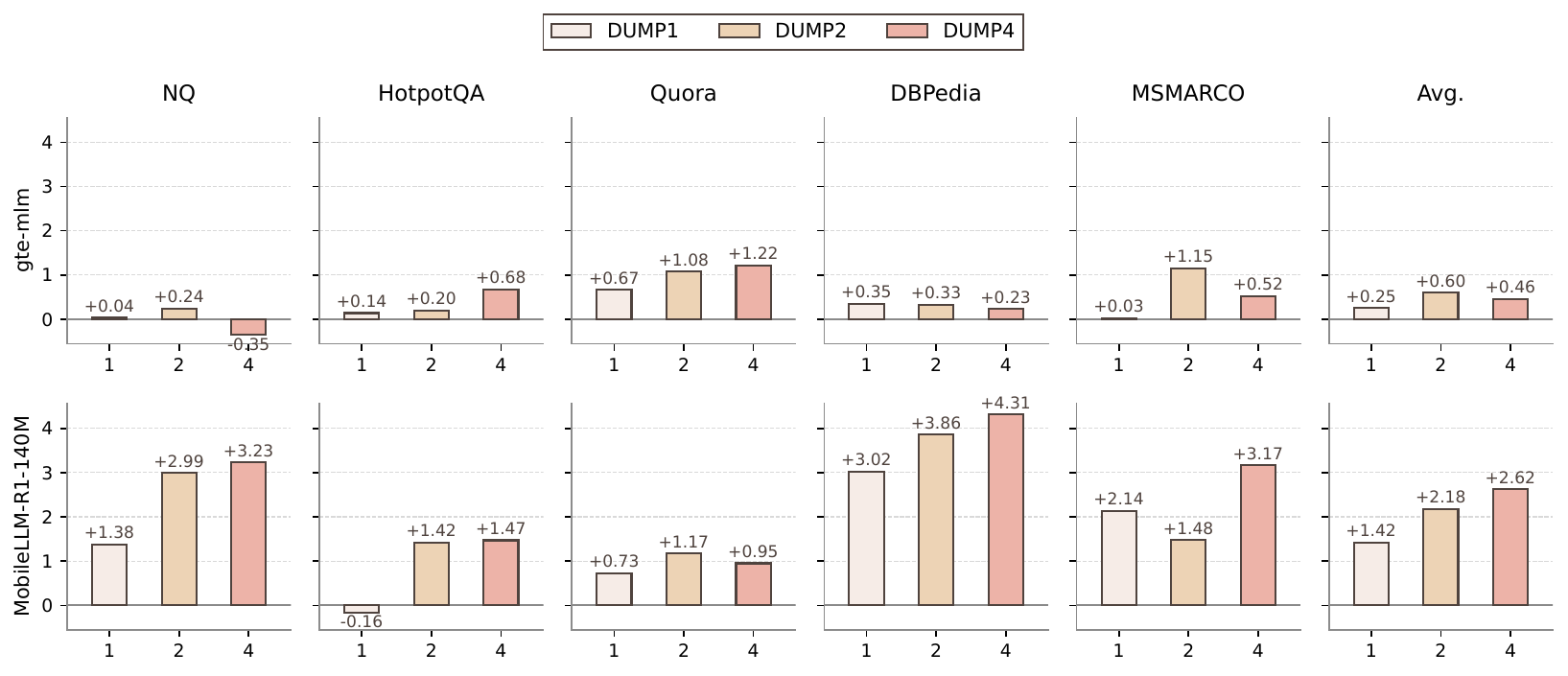}
    \caption{Effect of the number of \texttt{[DUMP]} tokens on retrieval performance. We compare one, two, and four \texttt{[DUMP]} tokens with gte-mlm-base as the encoder and MobileLLM-R1-140M as the decoder. Bars indicate performance changes relative to the corresponding baseline without \texttt{[DUMP]} tokens.}
    \label{fig:dump_token_num}
\end{figure*}

\smallskip
\noindent\textbf{Long-Context Retrieval Results.}
Table~\ref{tab:longembed} presents the LongEmbed evaluation results, where inputs are substantially longer than in standard retrieval benchmarks. This setting poses a stronger compression challenge because the final pooled embedding must encode information from a larger number of contextualized token representations. Our method consistently improves over the base models; for example, gte-en-mlm improves from 0.4797 to 0.5024, and ettin-decoder improves from 0.3330 to 0.3498. These gains suggest that trainable prefix tokens remain effective under long-context encoding, where residual token-level information can have a larger influence on the final pooled representation. This result is particularly meaningful because long-context inputs increase the risk that useful local evidence is weakened during pooling. The consistent gains suggest that trainable prefix tokens help the pooled representation preserve retrieval-relevant information even when the input contains many competing contextual signals.

\section{Analysis}
\subsection{Effect of the number of \texttt{[DUMP]} tokens.}
Figure~\ref{fig:dump_token_num} compares retrieval performance when varying the number of \texttt{[DUMP]} tokens, with $K \in \{1,2,4\}$. Adding these tokens generally improves performance over the baseline, in most cases across both models, but the gains are not strictly monotonic. For gte-mlm, using two tokens yields the best average improvement, whereas using four tokens slightly hurts NQ despite improving most other datasets. For MobileLLM-R1, using four tokens achieves the largest average gain, showing a clearer benefit from increasing the number of prefix tokens. Although the best token count varies across models and datasets, using \texttt{[DUMP]} tokens is generally more effective than not using them. These results suggest that the number of prefix tokens acts as a tunable hyperparameter. While additional tokens can improve retrieval performance, increasing the token count does not consistently lead to better results across all settings.

\begin{figure}[t!]
    \centering
    \includegraphics[width=\columnwidth]{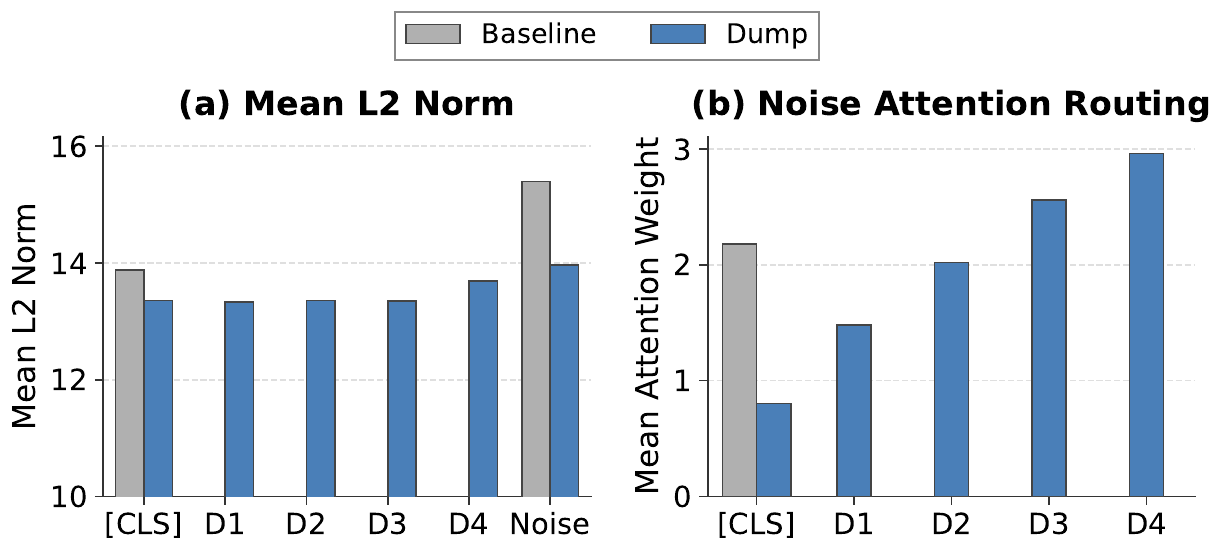}
    \caption{Token-level analysis of the \texttt{[DUMP]} mechanism using gte-en-mlm-base. Panel (a) compares token-group activation scales measured by the mean L2 norm. Panel (b) visualizes the attention distribution from low-information tokens to \texttt{[CLS]} and D1--D4, with D1--D4 ordered by prefix position.}
    \label{fig:dump_analysis}
\end{figure}

\subsection{The Role of Dump Tokens}
\paragraph{Setup}
To understand how \texttt{[DUMP]} tokens affect internal representations, we analyze token-level representations and attention routing using gte-en-mlm-base on LongEmbed NQ. In the \texttt{[DUMP]} setting, the model is trained with four prepended \texttt{[DUMP]} tokens, denoted as D1--D4. We group tokens into the sequence representation token (\texttt{[CLS]}), the learned prefix tokens (D1--D4), and low-information tokens, identified using NLTK English stopwords. Motivated by prior work showing that Transformer artifacts can be associated with unusually high-norm token representations~\cite{register}, Panel (a) measures the mean L2 norm of final-layer hidden states to compare activation scales across token groups. Panel (b) measures last-layer self-attention from low-information tokens to \texttt{[CLS]} and D1--D4, averaged over attention heads, and reports the weights after $10^5$ scaling for readability.

\paragraph{Results} Figure~\ref{fig:dump_analysis} summarizes the token-level analysis. As shown in Panel (a), the baseline model exhibits relatively high activation norms for low-information tokens, suggesting that these tokens can carry disproportionately large internal activations. With \texttt{[DUMP]} tokens, the L2 norm of low-information tokens is reduced, while the learned prefixes (D1--D4) form a stable activation range close to the \texttt{[CLS]} token. Panel (b) further shows that attention from low-information tokens is redistributed away from \texttt{[CLS]} and toward the learned prefixes. These results suggest that \texttt{[DUMP]} tokens provide alternative prefix positions to which attention from low-information tokens is redirected, thereby reducing their direct influence on the sequence representation.
\section{Conclusion}
We investigated whether trainable \texttt{[DUMP]} tokens can reduce representational interference in embedding models when prepended to the input text. Our approach preserves the original model architecture, pooling function, and retrieval objective, while allowing the added prefix tokens to be learned through the same contrastive training process. Across retrieval and long-context benchmarks, our method consistently improves over finetuned baselines, showing that learnable input prefixes can enhance retrieval representations without additional supervision or architectural changes. Beyond overall retrieval gains, our analysis indicates that \texttt{[DUMP]} tokens can change how low-information token signals are represented and attended to during encoding, providing evidence that the improvement is linked to internal representation behavior rather than only external performance differences. Overall, these findings suggest that trainable prefix tokens offer a simple and effective way to improve retrieval embeddings while preserving the underlying model architecture.

\section*{GenAI Usage Disclosure}
We used GenAI tools only for grammar checking and language polishing, not for method development, experiments, or result analysis.

\begin{acks}
This research was supported by Basic Science Research Program through the National Research Foundation of Korea(NRF) funded by the Ministry of Education(NRF-2021R1A6A1A03045425). This work was supported by Institute for Information \& communications Technology Promotion(IITP) grant funded by the Korea government(MSIT) (RS-2024-00398115, Research on the reliability and coherence of outcomes produced by Generative AI). This research was supported by Culture, Sports and Tourism R\&D Program through the Korea Creative Content Agency grant funded by the Ministry of Culture, Sports and Tourism in 2026 (Project Name: Development of an AI Agent Integrating Korean Language Knowledge for Personalized Language Consultation Services, Project Number: RS-2026-25506607, Contribution Rate: 30\%). This work was supported by the Commercialization Promotion Agency for R\&D Outcomes(COMPA) grant funded by the Korea government(Ministry of Science and ICT)(2710096072).
\end{acks}

\bibliographystyle{ACM-Reference-Format}
\bibliography{sample-base}

\end{document}